\documentclass{IEEEtran4PSCC}
\ifCLASSINFOpdf
   \usepackage[pdftex]{graphicx}
\else
   \usepackage[dvips]{graphicx}
\fi
\usepackage{booktabs}
\usepackage{dingbat}
\usepackage[ruled,vlined,shortend, linesnumbered]{algorithm2e} % algorithm
\usepackage[]{algorithmic} % algorithm

\usepackage[cmex10]{amsmath}
\usepackage{bm}

\usepackage{hyperref}

\usepackage{xcolor}

\makeatletter
\let\old@ps@headings\ps@headings
\let\old@ps@IEEEtitlepagestyle\ps@IEEEtitlepagestyle
\def\psccfooter#1{%
    \def\ps@headings{%
        \old@ps@headings%
        \def\@oddfoot{\strut\hfill#1\hfill\strut}%
        \def\@evenfoot{\strut\hfill#1\hfill\strut}%
    }%
    \def\ps@IEEEtitlepagestyle{%
        \old@ps@IEEEtitlepagestyle%
        \def\@oddfoot{\strut\hfill#1\hfill\strut}%
        \def\@evenfoot{\strut\hfill#1\hfill\strut}%
    }%
    \ps@headings%
}
\makeatother

\psccfooter{%
        \parbox{\textwidth}{\hrulefill \\ \small{23rd Power Systems Computation Conference} \hfill \begin{minipage}{0.2\textwidth}\centering \vspace*{4pt} \includegraphics[scale=0.06]{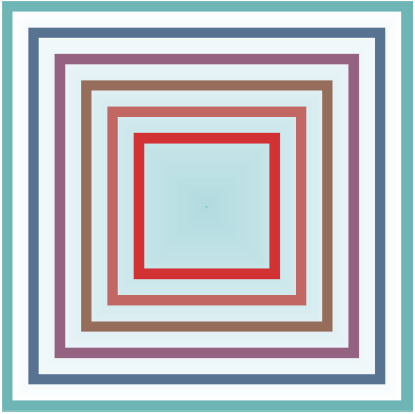}\\\small{PSCC 2024} \end{minipage} \hfill \small{Paris, France --- June 4 -- 7, 2024}}%
}

\begin{document}
\title{Contingency Analysis with Warm Starter using
Probabilistic Graphical Model}

%% To specify the authors when (number of affiliations <= 2)
\author{
\IEEEauthorblockN{Shimiao Li\\}
\IEEEauthorblockA{
%Department of Electrical and Computer Engineering, \\
Carnegie Mellon University,\\
Pittsburgh, USA\\
shimiaol@andrew.cmu.edu}
\and
\IEEEauthorblockN{Amritanshu Pandey}
\IEEEauthorblockA{University of Vermont\\
Burlington, USA\\
amritanshu.pandey@uvm.edu }
\and
\IEEEauthorblockN{Larry Pileggi}
\IEEEauthorblockA{
%Department of Electrical and Computer Engineering, \\
Carnegie Mellon University,\\
Pittsburgh, USA\\
pileggi@andrew.cmu.edu}
}

%% To specify the authors when (number of affiliations > 2)
% \author{\IEEEauthorblockN{Author n.1\IEEEauthorrefmark{1},
% Author n.2\IEEEauthorrefmark{2},
% Author n.3\IEEEauthorrefmark{3}, 
% Author n.4\IEEEauthorrefmark{3} and
% Author n.5\IEEEauthorrefmark{4}}
% \IEEEauthorblockA{\IEEEauthorrefmark{1} Department Name of Organization A\\
% Name of the organization A,
% Address A\\ Emails if wanted}
% \IEEEauthorblockA{\IEEEauthorrefmark{2} Department Name of Organization B\\
% Name of the organization B,
% Address B\\ Emails if wanted}
% \IEEEauthorblockA{\IEEEauthorrefmark{3} Department Name of Organization C\\
% Name of the organization C,
% Address C\\ Emails if wanted}
% \IEEEauthorblockA{\IEEEauthorrefmark{4}Department Name of Organization D\\
% Name of the organization D,
% Address D\\ Emails if wanted}
% }

\maketitle

\begin{abstract}
Cyberthreats are an increasingly common risk to the power grid and can thwart secure grid operations. We propose to extend contingency analysis to include cyberthreat evaluations. However, unlike the traditional N-1 or N-2 contingencies, cyberthreats (e.g., MadIoT) require simulating hard-to-solve N-k (with k $>>$ 2) contingencies in a practical amount of time. Purely physics-based power flow solvers, while being accurate, are slow and may not solve N-k contingencies in a timely manner, whereas the emerging data-driven alternatives are fast but not sufficiently generalizable, interpretable, and scalable. To address these challenges, we propose a novel conditional Gaussian Random Field-based data-driven method that performs fast and accurate evaluation of cyberthreats. It achieves speedup of contingency analysis by warm-starting simulations, i.e., improving starting points, for the physical solvers. To improve the physical interpretability and generalizability, the proposed method incorporates domain knowledge by considering the graphical nature of the grid topology. To improve scalability, the method applies physics-informed regularization that reduces model complexity. Experiments validate that simulating MadIoT-induced attacks with our warm starter becomes approximately 5x faster on a realistic 2000-bus system.
%requires, on average, 5x fewer iterations for a realistic 2000-bus system.
\end{abstract}

\begin{IEEEkeywords}
contingency analysis, cyber attack, Gaussian random field, power flow, warm start
\end{IEEEkeywords}

% Use this to place sponsorships
\thanksto{\noindent Accepted by the 23rd Power Systems Computation Conference (PSCC 2024).}

\section{Introduction}
\label{sec:Introduction}
The power grid is becoming increasingly challenging to operate. Climate change is accelerating the frequency of natural disasters, causing a higher rate of equipment failure, and new-age grid resources such as renewables and distributed energy resources are causing more uncertainty in supply. With the growing uncertainties and threats, today's grid operators use contingency analysis (CA) to identify the grid's vulnerability against them.

CA \cite{contingency-analysis} is a simulation module located within the grid control centers' Energy Management System (EMS). Grid operators (and planners) use it to perform a set of "what if" power flow simulations to preemptively evaluate the impacts of selected disturbances and outages on the power grid health indicators (e.g., voltages and line flow). {\color{black} If the output from CA indicates that the grid is operating outside of its operational limits}, the decision-makers can maintain reliability by taking {\color{black}remedial action}, for example, redispatching generation via a security-constrained optimal power flow (SCOPF). Today in operations, real-time CA is run every 5-30 minutes (e.g., Electric Reliability Council of Texas (ERCOT) which is the independent system operator in Texas runs it every 5 min \cite{ercot-rtca-5min} and North American Electric Reliability Corporation (NERC): requires that CA runs at least once every 30 min). Since it is computationally prohibitive to simulate all combinations of component failures in a limited time, the operators only include a predefined set of N-1 (loss of one component) contingencies in CA. These contingencies correspond to failures due to mechanical issues or natural disasters.

Today's CA methods may not suffice when facing a modern grid disturbance in the form of cyberthreat \cite{grid-security-Vyas}. Unlike N-1 contingencies, cyber threat can cause outages and malicious changes simultaneously at numerous locations, which represents N-k contingencies with $k>>1$.
%While the CA addresses challenges with loss of major equipment during mechanical failure and inclement weather, a modern grid disturbance in the form of cyberthreat is emerging where the today's CA may not suffice. 
Recent years' literature has documented many cyber attacks {\color{black}that target multiple locations}. Some describe a brute-force attack on many critical control devices (e.g. toggling main breakers of generators), causing blackout for days \cite{ukrain2015}\cite{lee2017crashoverride}; some describe hacking into a large set of grid-edge devices (e.g., Internet of Things-IoT devices) to deny or degrade electric service \cite{madiot}, and some describe disrupting the confidentiality and integrity of power grid data through the modification of many data values \cite{fdia-review}. 

With the emergence of these cyberthreats, the operators should preemptively secure the grid against them. 
%Protecting the grid operation from mechanical failures and natural events may not suffice. 
As such, in the future, CA should also include cyber events within its predefined set of contingencies. Simulating the power flow impact of these cyber events can inform the grid operators of the grid’s susceptibility to these attacks and also filter out potential incidents where preventive actions are needed. 

In this paper, we focus on a popular type of cyberthreat that can cause N-k contingencies: the MadIoT attack, also called BlackIoT attack, which has emerged due to the proliferation of the IoT devices (i.e., grid edge devices). This theoretical attack model succeeds by hacking into the high-wattage IoT-controlled load devices and adversely changing their load demand to disrupt the grid operation.  Attack instances of the MadIoT threat represent N-k events due to the concurrent load manipulations at many locations.  {\color{black} As this paper extends the functionality of contingency analysis unit  to also simulate such modern cyberattacks, we implicitly extended the definition of contingency to include the manipulation (which might not be a traditional failure) of devices that might make the system unstable.}

Unfortunately, the computational and analytical techniques in the EMS today may not solve CA robustly and fast enough for a large set of N-k contingencies events in an allocated amount of time, like within $\sim 5-30$ min for real-time CA. The main reason is that while fast evaluation of N-1 contingencies is possible due to the close proximity of post-contingency solution to the pre-contingency solution, these characteristics no longer hold true for N-k contingencies. Severe N-k contingencies can cause a big shift in the grid states, making it difficult to find \textit{sufficiently good} initial conditions that can lead to \textit{accurate} and \textit{fast} simulation.

%Various techniques can be applied to overcome challenges and quickly simulate N-k contingency events. One option is to develop contingency screening \cite{sugar-lodf} strategies that can significantly reduce the number of contingencies to simulate. 

{\color{black} Learning-based techniques, which can develop high-performance function approximators, provide an efficient way to address this challenge. Learning from some existing (simulated) N-k contingency data enables approximating the mapping from an original system to its state after a given disturbance. 
The goal of this paper is to develop such an approximator which serves as a warm starter that provides \textit{good and fast} initialization to CA and enables it to converge in fewer iterations.} Specifically, given a pre-contingency power grid and corresponding contingency information, the warm starter predicts the post-contingency bus voltages. The prediction is then used as the initial point for simulating the contingency.

Developing a data-driven learning-based method that is practical for grid-specific applications is important. General machine learning (ML) tools for physical systems exhibit the drawbacks of requiring a large amount of training data and outputting non-physical solutions. As an improvement, many current works of grid simulation and optimization have explored the use of physics, i.e., domain knowledge or domain expertise, towards developing physics-informed machine learning (ML) approaches \cite{rl-topology}\cite{rl-multiagent-v}-\cite{rl-multiagent-shunt}\cite{ppf-dnn}\cite{pf-dnn-topo}\cite{gnn-pf}\cite{dcopf-dnn}\cite{acopf-dnn-donti}\cite{gnn-acopf}\cite{gnn-acopf-warm}\cite{unrolled-se-2018}\cite{unrolled-se}\cite{unrolled-gnu-se}. However, we have found that many methods still lack sufficient generalization, interpretability and scalability. Section II-A provides an overview of these current efforts and their limitations.

%To develop the data-driven warm-starter, we focus on generalizability, physical interpretability and scalability. Current state-of-the-art methods do not satisfy these attributes. General machine learning (ML) tools while being fast may output non-physical solutions when applied to physical systems. Further, they may exhibit high variance and may not apply to extreme grid scenarios (i.e., low generalizability). As an improvement, many current works [xx]-[xx] of grid simulation and optimization have explored the use of \textit{physics}, i.e., domain knowledge or domain expertise, to improve physical interpretability and generalizability. However, we find that these methods still lack sufficient generalization, interpretability and scalability. Section \ref{sec:ML survey} provides an overview of these current efforts and their limitations. 

To address the limitations in the current physics-informed ML methods while developing a data-driven warm starter, this paper builds a probabilistic graphical model where the conditional joint distribution is factorized into a pairwise form. The potential functions (which are components of the factorization) are then defined in the form of Gaussian functions, giving rise to a conditional Gaussian Random Field (GRF) to model the conditional joint distribution. Neural networks are used to map from local inputs to the unknown parameters in local Gaussian potential functions, the training of which is based on a maximization of the conditional likelihood.
Such an integration of GRF with neural networks aims at improved 1) model generalizability by incorporating topology changes in the grid into the method by using architecture-constrained graphical models, 2) physical interpretability, since the inference model of our Gaussian Random Field has been found to form a linear proxy of the power system, and 3) trainability and scalability by using a graphical model with physics-informed regularization techniques (e.g., parameter sharing). The results show that on a 2000-bus system, the proposed warm starter enabled contingency simulation achieves 5x faster convergence than the traditional initialization methods. 

\textbf{Reproducibility:} our code is publicly available at \url{https://github.com/ohCindy/GridWarm.git}.

\section{Related Work} \label{sec:relatedwork}
\subsection{Physics-informed ML for power grids: literature review}\label{sec:ML survey}

Many prior works have included domain-knowledge in their methods to address the problem of missing \textit{physics} in generic ML tools. These methods collectively fall under physics-informed ML paradigm for power grid operation, control, and planning and can be broadly categorized into following categories:

\subsubsection{Reducing search space}
In these methods, domain knowledge is used to narrow down the search space of parameters and/or solutions. For example, \cite{rl-topology} designed a grid topology controller which combines reinforcement learning (RL) Q-values with power grid simulation to perform a 
\textit{physics}-guided action exploration, as an alternative to the traditional epsilon-greedy search strategy. Works in \cite{rl-multiagent-v}-\cite{rl-multiagent-shunt} studied the multi-agent RL-based power grid control. In these approaches the power grid is partitioned into  controllable sub-regions based on domain knowledge (i.e., electrical characteristics) to reduce the high-dimensional continuous action space into lower-dimension sub-spaces which are easier to handle.

\subsubsection{Enforcing system constraints and technical limits}

Many recent works apply deep learning to power grid analysis problems. These include but are not limited to power flow (PF) \cite{ppf-dnn}\cite{pf-dnn-topo}\cite{gnn-pf}, DC optimal power flow (DCOPF)\cite{dcopf-dnn}, ACOPF \cite{acopf-dnn-donti}\cite{gnn-acopf}\cite{gnn-acopf-warm} and state estimation (SE) \cite{unrolled-se-2018}\cite{unrolled-se}\cite{unrolled-gnu-se}. These methods are generally based on (supervised) learning of an input-to-solution mapping using historical system operational data or synthetic data. A type of method amongst these works is unrolled neural networks \cite{unrolled-se-2018}\cite{unrolled-se}\cite{unrolled-gnu-se} whose layers mimic the iterative updates to solve SE problems using first-order optimization methods (i.e., gradient descent methods), based on quadratic approximations of the original problem. 
Many methods learn the 'one-step' mapping function. 
Among these, some use deep neural network (DNN) architectures \cite{ppf-dnn}\cite{pf-dnn-topo}\cite{acopf-dnn-donti}\cite{dcopf-dnn} to learn high-dimensional input-output mappings, some use recurrent neural nets (RNNs) \cite{rnn-dse} to capture grid dynamics, and others apply graph neural networks (GNN) \cite{gnn-pf}\cite{gnn-acopf}\cite{gnn-se} to capture the exact topological structure of power grid.

To promote \textit{physical} feasibility of the solution, many works impose equality or inequality system constraints by i) encoding hard constraints inside NN layers (e.g. using sigmoid layer to encode technical limits of upper and lower bounds), ii) applying prior on the NN architecture (e.g., Hamiltonian \cite{hamiltonianNN} and Lagrangian
neural networks \cite{lagrangianNN}), iii) augmenting the objective function with penalty terms in a supervised \cite{ppf-dnn} or unsupervised \cite{acopf-dnn-donti}\cite{pf-dnn-topo} way, iv) projecting outputs \cite{dcopf-dnn} to the feasible domain, or v) combining many different strategies. 
%In general, ML models with system characteristics taken into account are often named as \textit{model-based}, compared with a blind ML blackbox which is \textit{model-free}. 
In all these methods, incorporating (nonlinear) system constraints remains a challenge, even with state-of-the-art toolboxes \cite{neuromancer2022}, and most popular strategies lack rigorous guarantees of nonlinear constraint satisfaction. 

While these methods have advanced the state-of-the-art in physics-informed ML for power grid applications, critical limitations in terms of generalization, interpretation, and scalability exist. We discuss these further:

\textbf{Limited generalization:} Many existing methods do not adapt well to changing grid conditions. Take changes in network topology as an example. Many current works are built on non-graphical architectures without any topology-related inputs. These, once trained, only work for one fixed topology and cannot generalize to dynamic grid conditions. More recently, some works have begun to encode topology information. Graph model-based methods (e.g., GNN \cite{gnn-pf}\cite{gnn-acopf}\cite{gnn-se}) naturally impose topology as a hard constraint and thus can account for topology changes. Alternatively, work in \cite{pf-dnn-topo} encodes the topology information into the penalty term (as a soft constraint) through the admittance and adjacency matrix, and \cite{unrolled-gnu-se} accounts for topology in NN implicitly by applying a topology-based prior through a penalty term. While these methods lead to better topology adaptiveness, they also have some risks: the use of penalty terms \cite{pf-dnn-topo}\cite{unrolled-gnu-se} to embed topology information as a soft constraint can lead to limited precision; and, for problems (like OPF) where information needs to be exchanged between far-away graph locations, the use of GNNs requires carefully designed global context vectors to output predictions with global-level considerations. 
%Our paper designs a graph model based method that can exactly encode different topologies.

\textbf{Limited interpretability:} Despite that many ML models (NN, decision trees, K-nearest-neighbors) are universal approximators, interpretations of their functionality from a physically meaningful perspective are still very limited. The general field of \textit{model interpretability} \cite{interpretability} focuses on explaining \textit{how a model works}, to mitigate fears of the unknown.
Broadly, investigations of interpretability have been categorized into transparency (also called ad-hoc interpretations) and post-hoc interpretations. The former aims to elucidate the mechanism by which the ML \textit{blackbox} works before any training begins by considering the notions of \textit{simulatability} (Can a human work through the model from input to output, in reasonable time steps through every calculation required to
produce a prediction?), \textit{decomposability} (Can we attach an intuitive explanation to each part of a model: each input, parameter, and calculation?), and \textit{algorithmic transparency} (Does the learning algorithm itself confer guarantees on convergence, error surface even for unseen data/problems?). And \textit{post-hoc interpretation} aims to inspect a learned model after training by considering its natural language explanations, visualizations
of learned representations/models, or explanations of empirical
examples. However, none of these concepts in the field of ML model interpretability formally evaluates \textit{how a ML model makes predictions in a physically meaningful way} when it is used on an industrial system like the power grid.  Some recent works have explored the physical meaningfulness of their models from the power system perspective. Still, interpretations are made in conceptually different ways without uniform metric: Unrolled neural networks (which has been used as data-driven state estimation for power grid \cite{unrolled-gnu-se}\cite{unrolled-se-2018}) are more decomposable and interpretable in a way that the layers mimic the iterations in the physical solvers, yet these models \cite{unrolled-se}\cite{unrolled-se-2018}\cite{unrolled-gnu-se} mainly unroll first-order solvers instead of the second-order (Newton-Raphson) realistic solvers.
GNN-based models \cite{gnn-pf}\cite{gnn-acopf}\cite{gnn-se} naturally enable better interpretability in terms of representing the graph structure. 
Work in \cite{ppf-dnn} provides some interpretation of its DNN model for PF by matching the gradients with power sensitivities and subsequently accelerating the training by pruning out unimportant gradients. \cite{pf-dnn-topo} learns a weight matrix that can approximate the bus admittance matrix; however, with only limited precision. To summarize, due to the limited interpretability, ML models still have some opacity and blackbox-ness, when compared with the purely physics-based models (e.g., power flow equations).
%Our paper addresses this issue by defining a new concept \textit{physical interpretability}. 

\textbf{Scalability issues}:
In the case of large-scale systems, models (like DNNs) that learn the mapping from high-dimensional input-output pairs will inevitably require larger and deeper designs of model architecture and thereafter, massive data to learn such mappings. This can affect the practical use in real-world power grid analytics.
%This paper addresses these issues through a distributed scheme and the use of domain knowledge to regularize the model effectively.

\subsubsection{Extracting meaningful features or crafting an interpretable latent space} 
Many works exist in this class. For example, \cite{pmu-graph-temporal}\cite{pmu-graph-spatial} learned the latent representation of sensor data in a graph to capture temporal dependency  \cite{pmu-graph-temporal} or spatial sensor interactions \cite{pmu-graph-spatial}.
\cite{cascade-influence-model} applied the influence model to learn, for all edge pairs, the pairwise influence matrices, which are then used to predict line cascading outages. \cite{dynwatch} crafted a graph similarity measure from power sensitivity factors and detected anomalies in the context of topology changes by weighing historical data based on this similarity measure.

\subsection{MadIoT: IoT-based Power Grid Cyberthreat}\label{sec:madiot}

\begin{figure}[h]
	\centering
	\includegraphics[width=1.0
	\linewidth]{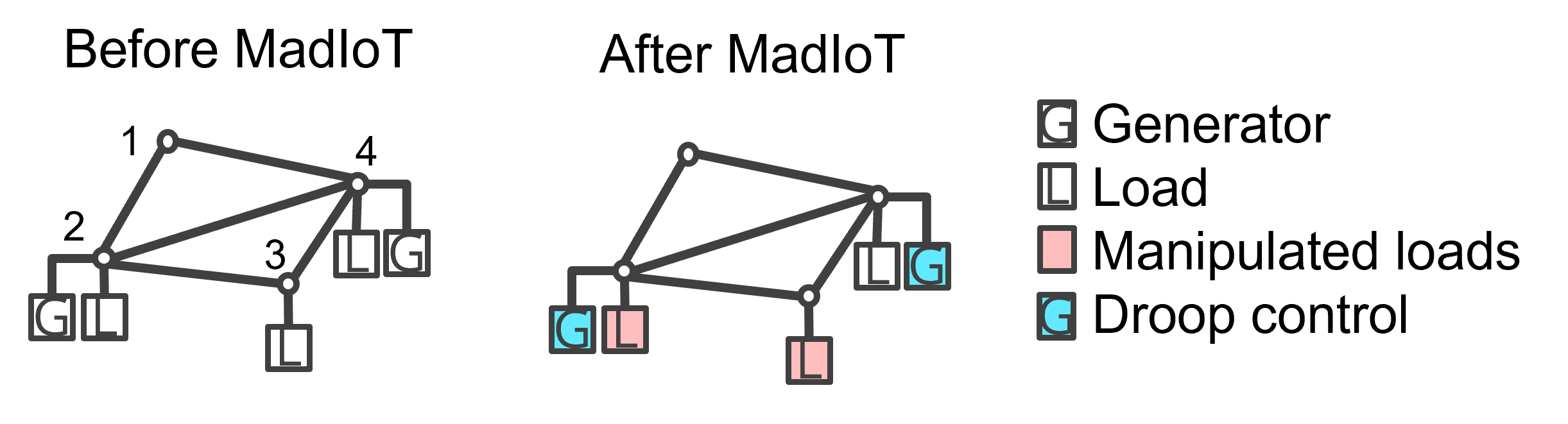}
	\caption[]{A toy instance of MadIoT: a subset of loads is manipulated.}
	\label{fig:madiot}
\end{figure}

The proliferation of IoT devices has raised concerns about IoT-induced cyber attacks. \cite{madiot} proposed a threat model, namely \textit{BlackIoT} or \textit{MadIoT}, where an attacker can manipulate the power demand by synchronously turning on/off or scaling up/down some high-wattage
IoT-controlled loads on the grid. This can lead to grid instability or grid collapse. Fig. \ref{fig:madiot} illustrates an attack instance of the MadIoT threat. To evaluate the impact of a postulated attack, \cite{madiot} uses a steady-state analysis of the power grid while considering the droop control, protective relaying, and thermal and voltage limits for various components.

% \subsection{Contingency analysis}\label{sec:pf and ca}
% Contingency Analysis (CA)\cite{contingency-analysis} is a 'what if' simulation used in both operation and planning to evaluate the line flow and voltage impacts of possible disturbances and outages. Today during operation, real-time CA is performed approximately every 5-30 minutes (5min in ERCOT\cite{ercot-rtca-5min}, at least 30min required by NERC). The CA simulates a predefined set of N-1 (loss of one element) contingencies. In near future, with potential likelihood of cyberattacks (e.g., MadIoT), the power grid will have to be evaluated under a new set of cyberattack-based contingencies for reliable and secure operation. Instead of  traditional N-1 contingencies, these contingencies will represent N-k events (simultaneous loss of x components).
% While fast evaluation of many N-1 contingencies is possible due to close proximity of final solutions to initial conditions, these characteristics no longer hold true for cyberattack-based N-k contingencies. Therefore, it may be a challenge to evaluate them in a limited time.

% To address this challenge, one option is to use effective contingency screening \cite{sugar-lodf} strategies that can significantly reduce the number of contingencies to simulate. Another option is to have a warm starter (i.e., improved initial conditions) that can speed up the convergence of each simulation. This paper proposes a method for the latter.

%\input{030framework}
\section{A novel warm starter} \label{sec:method}
\noindent \textbf{Goal}: This proposed warm starter will enable simulating a large-set of cyberthreat-driven contingencies $(C_{N-k})$ in a practical amount of time by improving the speed of individual simulation. 

Purely physics-based solvers can give accurate solutions but are slow when they have to solve a large number of \textit{hard-to-solve} N-k contingency events. In contrast, replacing physical solvers with purely data-driven techniques makes it fast but has severe limitations, as discussed in Section \ref{sec:ML survey}. In this paper, we propose a novel physics-informed ML model that can warm-start the physical solvers when simulating \textit{hard-to-solve} contingencies  $(C_{N-k})$. The warm starter will predict the post-contingency voltages supplied as initial conditions to the physical solver for fast convergence. The proposed model is designed to be \textbf{generalizable to topology change} by using a graphical model, and \textbf{physically interpretable} by forming a linear system proxy, and \textbf{scalable} by application of regularization techniques on the graphical model.

\subsection{Task definition and symbol notations} 
As shown in Fig. \ref{fig: graphical model}, given an input $\bm{x}$ which contains contingency information $c$ and (pre-contingency) system information $G$, a warm starter makes prediction $\bm{y}$ which is an estimate of the post-contingency bus voltages $\bm{v}^{post}$. The model is a function mapping, which is learned from training dataset  $Data=\{(\bm{x^{(j)},\bm{y^{(j)}}})\}$, where $(j)$ denotes the $j$-th sample. Table \ref{tab:notations} shows the symbols used in this paper.

%To warm start the contingency analysis on MadIoT (see Figure \ref{fig:madiot}), we aim to obtain a rough estimate of the post-contingency bus voltages $\bm{v}^{post}$, given  the contingency setting $c$, the pre-contingency case information $G$, which includes topology, power settings and the pre-contingency bus voltages $\bm{v}^{pre}$. To this end, we seek to train a model of mapping function, using the dataset  $Data=\{(\bm{x^{(j)},\bm{y^{(j)}}})\}$, where $\bm{x^{(j)}}$ denotes the input features of the $j$-th sample containing input feature about $G^{(j)},c^{(j)}$; $\bm{y^{(j)}}$ stores $\bm{v^{post(j)}}$ as the (ground truth) labels.

\setlength{\tabcolsep}{6pt}
\begin{table}[htbp]
\small
\centering
	\caption{Symbols and definitions \label{tab:notations}}
	\begin{tabular}{ @{}rl@{} }  
	\toprule
	\textbf{Symbol} & \textbf{Interpretation} \\ \midrule
		$G$    
     &case data before contingency\\
     & containing topology, generation, and load settings\\
     \midrule
     $\bm{v_i}$   & the voltage at bus $i$,
     $\bm{v_i}=[v_i^{real},v_i^{imag}]^T$\\
     \midrule
     $\bm{v}^{pre/post}$ & $\bm{v}^{pre/post}=[\bm{v}^{pre/post}_1,\bm{v}^{pre/post}_2,...,\bm{v}^{pre/post}_n]^T$ \\
     &the pre/post-contingency voltages at all buses\\
     \midrule
     $c$ & contingency setting $\textit{(type, location, parameter)}$\\
     &e.g. $(\textit{MadIoT}, [1,3], 150\%)$: increasing loads at bus 1\\
     & and 3 to $150\%$ of the original value via MadIoT attack.\\
     \midrule
     $i,n$ & bus/node index; total number of nodes\\
    %  $k,K$ & index of branch/edge; total number of branches/edges\\
     $(s,t)$& a branch/edge connecting node $s$ and node $t$\\
     $\mathcal{V},\mathcal{E}$ & set of all nodes and edges: $i\in\mathcal{V},\forall i; (s,t)\in\mathcal{E}$ \\
     $j,N$ & data sample index; total number of {\color{black} (training)} samples\\
     \midrule
     $(\bm{x},\bm{y})$ & a sample with feature $\bm{x}$ and output $\bm{y}$\\
     &$\bm{y}=[\bm{y_1,...,y_n}]^T=[\bm{v}^{post}_1,...,\bm{v}^{post}_n]^T$ \\
	\bottomrule
	\end{tabular} 
\end{table}

\subsection{Method Overview}

\begin{figure}[h]
	\centering
	\includegraphics[width=0.6\linewidth]{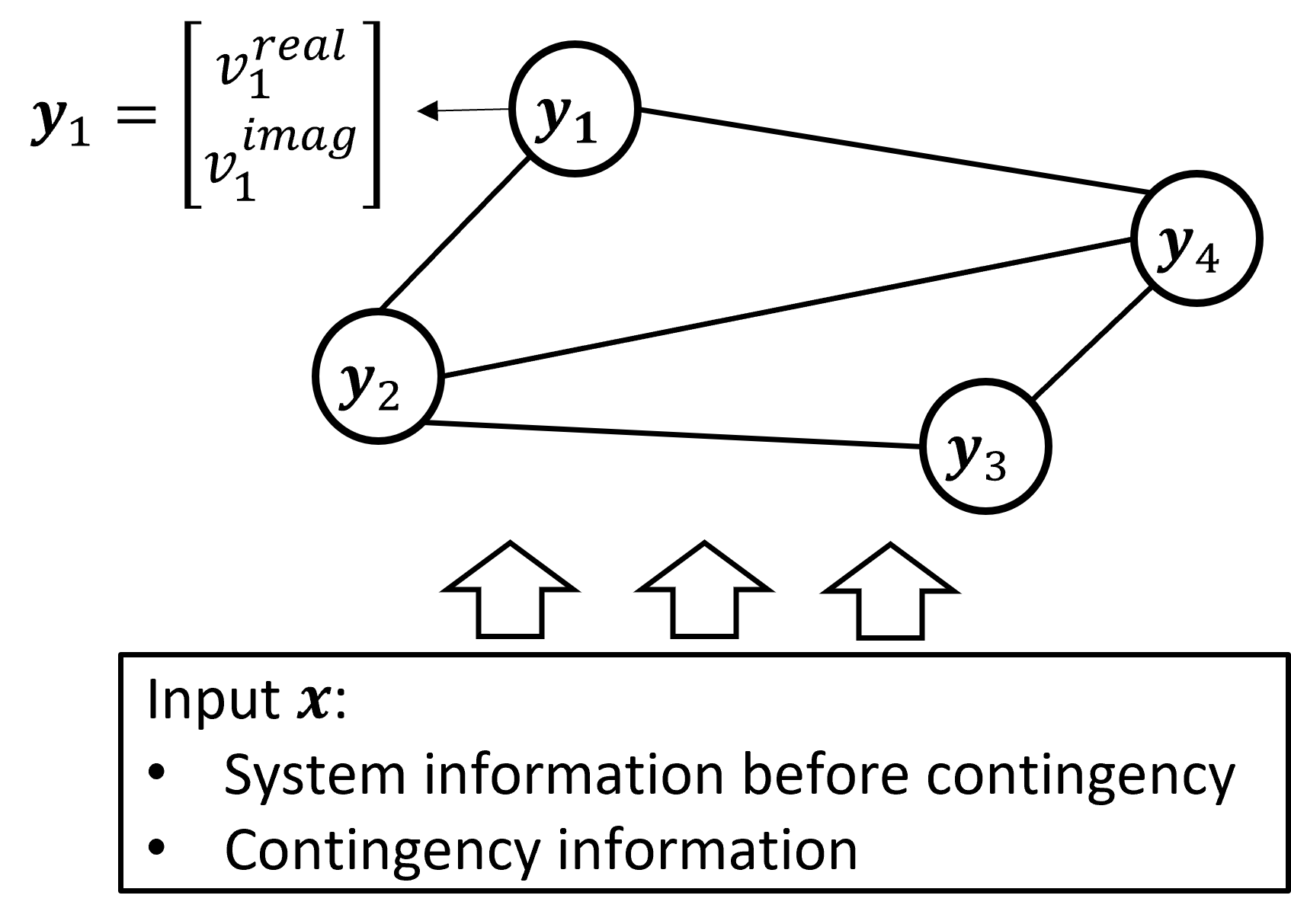}
	\caption[]{\color{black} A power grid can be naturally represented as a graphical model. Each node represents the bus voltage after contingency, each edge represents a branch status after contingency. Now conditioned on an original power grid $G$ and a contingency $c$ that happens on it, we want to know the bus voltages after contingency. }
	\label{fig: graphical model}
\end{figure}

Power grid can be naturally represented as a graph, as shown in Fig. \ref{fig: graphical model}.
Nodes and edges on the graph correspond to power grid buses and branches (lines and transformers), respectively. Each node represents a variable $\bm{y_i}$ which denotes voltage phasor at bus $i$, whereas each edge  represents a direct inter-dependency between adjacent nodes. {\color{black}Such  an undirected graphical model that models the dependency relationship among the random variables is called a Markov Random Field (MRF).}

%We can show that the graphic representation naturally models the conditional independence (CI) relationships on the power grid. And the local Markov property, Pairwise Markov Property and Global Markov Property hold true with real physical meanings, as Appendix \ref{appendix:markov properties} shows.
%As the power balance is satisfied on every bus, it is physically true that given the states on all its neighbors, the state of $\forall$ node $i$ is only dependent on the local network parameters (transformer and transmission line admittances), whereas conditionally independent from the rest of the nodes. This indicates that, for $\forall$ node $i$, the set of all neighboring nodes forms its Markov blanket, and thus the local Markov property holds with real-physical meaning. And one can easily show that global Markov property and pairwise Markov property also hold true with physical interpretations.

{\color{black}The use of MRF} enables a compact way of writing the conditional joint distribution and performing inference thereafter, using observed data. 
Specifically, when contingency happens, the joint distribution of variables $\bm{y}$ conditioned on input features $\bm{x}$ can be further factorized in a pairwise manner, {\color{black} leading to a pairwise Markov Random Field \cite{MRF}, as Fig.\ref{eq:pairwise parameterization} shows:}
\begin{align}
    &p(\bm{y}|\bm{x},\bm{\theta})=\frac{1}{Z(\bm{\theta},\bm{x})}\prod_{i=1}^n\psi_i(\bm{y_i}) \prod_{(s,t)\in\bm{E}}\psi_{st}(\bm{y_s},\bm{y_t})
    \label{eq:pairwise parameterization}
\end{align}
where $\bm{\theta}$ denotes model parameters that maps $\bm{x}$ to $\bm{y}$; $\psi_i(\bm{y_i}),\psi_{st}(\bm{y_s},\bm{y_t})$ are node and edge potentials conditioned on $\bm{\theta}$ and $\bm{x}$; and $Z(\bm{\theta},\bm{x})$ is called the partition function that normalizes the probability values such that they sum to the value of 1.

% In this conditional random field (CRF) modeling, 
% %Given a dataset of $\{(\bm{x}^{(j)},\bm{y}^{(j)})\}$,
% the potential functions are data-dependent, i.e., conditioned on input features. 
The factorization model in (\ref{eq:pairwise parameterization}) is inspired by pairwise continuous MRF \cite{MRF} and has an intuitive form: every edge potential encodes the mutual correlation between two adjacent nodes; both node and edge potentials represent the local contributions of nodes/edges to the joint distribution. In the task of contingency analysis, each potential function intuitively represents how the status of each bus and branch 'independently' impacts the bus voltages.
%There can be several ways to interpret why the post-contingency voltage variables are modeled as a probability distribution (see Appendix \ref{appendix:v distribution}). 

Given a training dataset of $N$ samples $\{(\bm{x}^{(j)},\bm{y}^{(j)})\}$, the training and inference can be described briefly as:
\begin{itemize}
    \item \textbf{Training:} With proper definition of the potential functions $\psi_i(\bm{y_i}|\bm{x,\theta})$, $\psi_{st}(\bm{y_s},\bm{y_t}|\bm{x,\theta})$ (see Section \ref{sec:cGRF} and \ref{sec: NN-node and NN-edge}), and the parameter $\bm{\theta}$ can be  learned by maximizing log-likelihood
   \begin{equation}
       \bm{\hat{\theta}}=arg\max_{\bm{\theta}} \sum_{j=1}^N log l(\theta)^{(j)}
   \end{equation}
   where $l(\theta)^{(j)}$ denotes the likelihood of the j-th sample:
   \begin{equation}
       l(\theta)^{(j)}=p(\bm{y}^{(j)}|\bm{x}^{(j)},\bm{\theta})
   \end{equation}
    \item \textbf{Inference:} For any new input $\bm{x}$, we make use of the estimated parameter $\hat{\bm{\theta}}$ to make a single-point prediction 
    \begin{equation}
        \bm{\hat{y}_{test}}=arg\max_{\bm{y}} p(\bm{y|x,\hat{\theta}})
        \label{eq: general inference}
    \end{equation}
\end{itemize}

The use of a probabilistic graphical setting naturally integrates the domain knowledge from grid topology into the method:  

\noindent \textbf{Domain knowledge of grid topology:} \textit{Power flow result is conditioned on the grid topology. Bus voltages of two adjacent buses connected and directed by a physical linkage (line or transformer) have direct interactions.
}

Each sample in this method can have its topology and each output is conditioned on its input topology. The following sections will discuss how the graphical model and domain knowledge enable an efficient and physically interpretable model design.

\subsection{Pairwise Conditional Gaussian Random Field} \label{sec:cGRF}
Upon representing the power grid and its contingency as a conditional pairwise MRF factorized in the form of (\ref{eq:pairwise parameterization}), we need to define the potential functions $\psi_i(\bm{y_i}),\psi_{st}(\bm{y_s},\bm{y_t})$. 

This paper builds a Gaussian random field which equivalently assumes that the output variable $(\bm{y})$ satisfies multivariate Gaussian distribution, i.e., $P(\bm{y}|\bm{x},\bm{\theta})$ is Gaussian. The justification and corresponding benefits of using Gaussian Random Field are:
\begin{itemize}
    \item partition function $Z(\bm{\theta},\bm{x})$ is easier to compute due to nice properties of Gaussian distribution. {\color{black} Specifically in the case of Gaussian, the normalization constraint can be computed easily by calculating matrix determinant $|\Lambda|$, whereas the use of other potential functions might lead  computation difficulties, potentially NP-hard \cite{MRF}.}
    \item high physical interpretability due to a physically meaningful inference model. We will discuss this later.
\end{itemize}

\noindent The potential functions for Gaussian random field \cite{MRF} are defined as follows:
\begin{align}
    &\psi_i(\bm{y_i})=exp(-\frac{1}{2}\bm{y_i}^T\bm{\Lambda_{i}}\bm{y_i}+\bm{\eta_i}^T\bm{y_i})\label{eq:gaussian_node_potential}\\
    &\psi_{st}(\bm{y_s},\bm{y_t})=exp(-\frac{1}{2}\bm{y_s}^T\bm{\Lambda_{st}}\bm{y_t})
    \label{eq:gaussian_edge_potential}
\end{align}
where $\bm{\Lambda_{i}}$ is a matrix and $\bm{\Lambda_{st}}$ is a vector.
By plugging \eqref{eq:gaussian_node_potential} and \eqref{eq:gaussian_edge_potential} into \eqref{eq:pairwise parameterization}, we have:
\begin{equation}
    p(\bm{y}|\bm{x},\bm{\theta})\propto exp(\bm{\eta}^T\bm{y}-\frac{1}{2}\bm{y}^T\Lambda\bm{y})
    \label{eq:gaussianMRF}
\end{equation}
where $\bm{\Lambda_{i}}$ and $\bm{\Lambda_{st}}$ parameters are the building blocks of matrix $\bm{\Lambda}$, and $\bm{\eta}$ is a column vector composed of all $\bm{\eta_i}$.
To further illustrate, consider a post-contingency grid structure in Fig. \ref{fig: graphical model}. The $\bm{\eta}$ and $\bm{\Lambda}$ variables for this grid structure will be shown later (in Fig. \ref{fig: training process} where the $\bm{0}$ blocks in $\bm{\Lambda}$ matrix are structural zeros representing no \textit{active} edges at the corresponding locations).

In the model (\ref{eq:gaussianMRF}),   both $\bm{\eta}$ and $\bm{\Lambda}$ are functions of $\bm{x},\bm{\theta}$, i.e.,
\begin{equation}
    {\bm{\eta}}=f_\eta(\bm{x},\bm{{\theta_{\eta}}}), {\bm{\Lambda}}=f_\Lambda(\bm{x},\bm{{\theta_{\Lambda}}})
\end{equation}
and $P(\bm{y}|\bm{x},\bm{\theta})$ takes an equivalent form of a multivariate Gaussian distribution $N(\bm{\mu},\bm{\Sigma})$ ($\bm{\mu}$ is the mean and $\bm{\Sigma}$ is the covariance matrix) with
\begin{equation}
\bm{\eta}=\bm{\Lambda}\bm{\mu}, \bm{\Lambda}=\bm{\Sigma}^{-1}    
\end{equation}

Now based on these defined models, we seek to learn the parameter $\bm{\theta}$ through maximum likelihood estimation (MLE). The log-likelihood of each data sample can be calculated by:
\begin{equation}
    l(\bm{\theta})= logP(\bm{y}|\bm{x},\bm{\theta})= -\frac{1}{2}\bm{y}^T\bm{\Lambda}\bm{y}+\bm{\eta}^T\bm{y}-log Z(\bm{\theta},\bm{x}) \label{eq: loglikelihood}
\end{equation}
and the MLE can be written equivalently as an optimization problem that minimizes the negative log-likelihood loss on the data set of $N$ training samples:
\begin{equation}
    \min_{\bm{\theta}} -\sum_{j=1}^{N}l(\bm{\theta})^{(j)}
    \label{eq:MLE simplified}
\end{equation}

\textbf{Inference and Interpretation:}
Upon obtaining the solution of $\hat{\bm{\theta}}=[\hat{\bm{\theta_{\eta}}},\hat{\bm{\theta_{\Lambda}}}]^T$, parameters $\hat{\bm{\Lambda}}=f_\Lambda(\bm{x_{test}},\bm{\hat{\theta}})$, $\hat{\bm{\eta}}=f_\eta(\bm{x_{test}},\bm{\hat{\theta}})$ can be estimated thereafter. Then for any test contingency sample $\bm{x_{test}}$, the inference model in (\ref{eq: general inference}) is equivalent to solving $\bm{\hat{y}_{test}}$ by:
\begin{align}
    \hat{\bm{\Lambda}}\bm{\hat{y}_{test}}=\hat{\bm{\eta}}
    \label{eq: linear approx}
\end{align}

Notably, the model in (\ref{eq: linear approx}) can be seen as a \textbf{linear system proxy} of the post-contingency grid, providing a physical interpretation of the method. $\bm{\Lambda}$ is a sparse matrix with a structure similar to the bus admittance matrix where the zero entries are 'structural zeros' representing no branch connecting buses. $\bm{\eta}$ behaves like the net injection to the network. 

\subsection{NN-node and NN-edge}\label{sec: NN-node and NN-edge} 

%this section can be simplified by starting with lambda_ii, lambda_st, and \eta

Finally to implement the model, %formulated as optimizing the log-likelihood, or equivalently minimizing the negative log-likelihood loss, 
we need to specify the functions of  $f_\eta(\bm{x},\bm{{\theta_{\eta}}}),f_\Lambda(\bm{x},\bm{{\theta_{\Lambda}}})$. 
%In the way we defined the potential functions, despite that the size of $\bm{\Lambda,\eta}$ increases with grid size, we can take advantage of the sparsity of $\bm{\Lambda}$ so that only some edge-wise parameters $\bm{\Lambda_{st}}$ and node-wise parameters $\bm{\Lambda_{i}},\bm{\eta_i}$ have to be learned.
Taking advantage of the sparsity of $\bm{\Lambda}$, the task here is to learn a function mapping from input $\bm{x}$ to only some edge-wise parameters $\bm{\Lambda_{st}}$ and node-wise parameters $\bm{\Lambda_{i}},\bm{\eta_i}$. Yet the number of $\bm{\Lambda_{i}},\bm{\Lambda_{st}}, \bm{\eta_i}$ parameters still increases with grid size, meaning that the input and output size of the model will explode for a large-scale system, requiring a much more complicated model to learn a high-dimensional input-output map. 

\begin{figure}[h]
	\centering
	\includegraphics[width=1.0\linewidth]{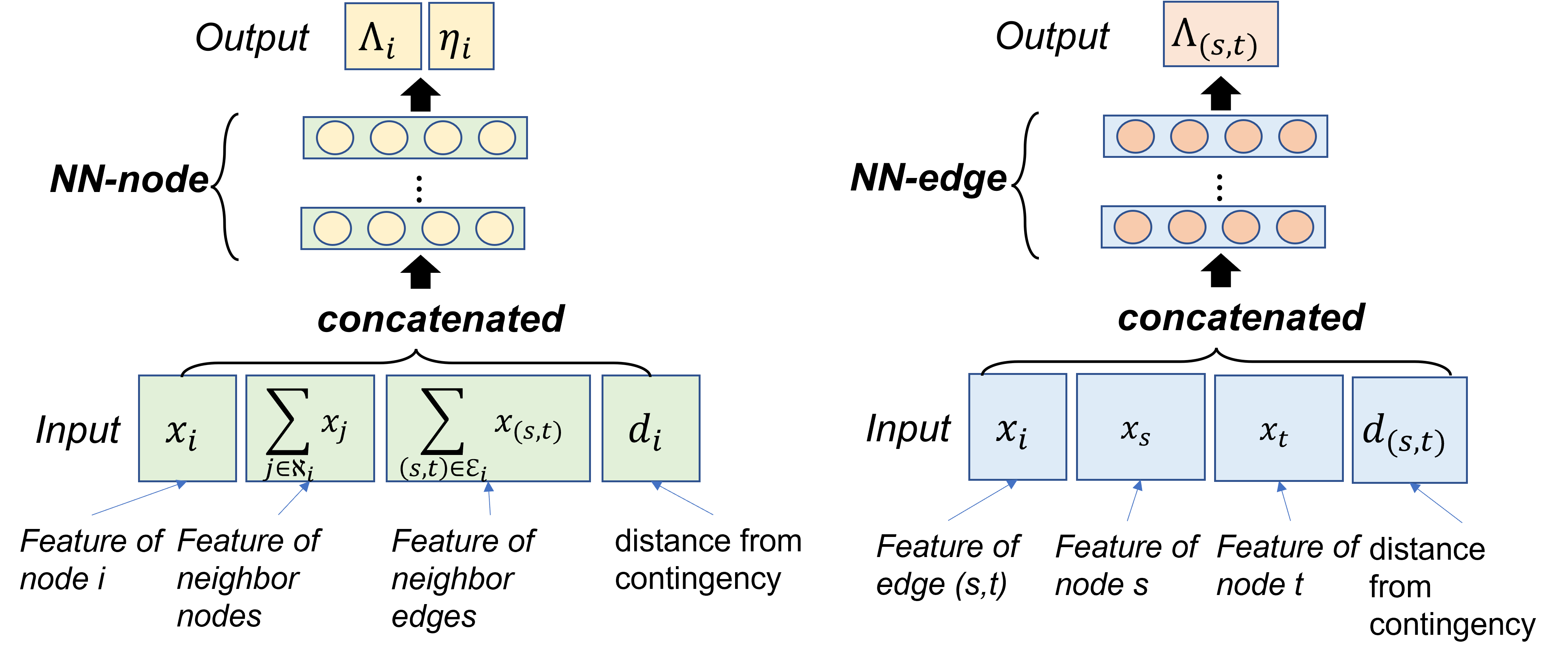}
	\caption[]{each node has a \textit{NN-node} and each edge has a \textit{NN-edge}, to map the input features to the post-contingency system characteristics.}
	\label{fig: NN-node and NN-edge}
\end{figure}

To efficiently reduce the model size, this paper implements the mapping functions using local Neural Networks: each node has a \textit{NN-node} to predict $\bm{\Lambda_{i}},\bm{\eta_i}$ using local inputs; each edge has a \textit{NN-edge} to output $\bm{\Lambda_{st}}$ in a similar way, as shown in Fig. \ref{fig: NN-node and NN-edge}. This is inspired by our interpretation that $\bm{\Lambda}$ is a proxy of the bus admittance matrix whose elements represent some local system characteristics regarding each node and edge.

Meanwhile, to effectively learn the mapping, we must answer the following question: \textbf{how to select the input features to the NN models optimally?} We apply \textit{domain knowledge} to design the input space that feeds most relevant features into the model:\\

\noindent \textbf{Domain knowledge of decisive features:} \textit{The impact of contingency depends heavily on the importance of contingency components which can be quantified by the amount of its generation, load or power delivery.
}

\noindent \textbf{Domain knowledge of Taylor Expansion on system physics:}\textit{ Let $v=h(G)$ denote any power flow simulation that maps the case information to the voltage profile solution. By Taylor Expansion, the post-contingency voltage can be expressed as a function depending on pre-contingency system $G_{pre}$ and the system change $\Delta G$ caused by contingency:
$$\bm{v^{post}} = h(G_{pre}) + h' (G_{pre})\Delta G + \frac{1}{2}h'' (G_{pre})\Delta G^2 + ...$$
}

Therefore, the key features of the pre-contingency system $(G_{pre})$ and system change $(\Delta G)$ are selected as node features to feed into the NN mappings, and include:
\begin{itemize}
    \item node feature $\bm{x}_i$: real and imaginary voltages ($v_i^{real}, v_i^{imag}$), power and current injections ($ P_i, Q_i, I_i^{real}, I_i^{imag}$), and shunt injections ($Q_{shunt,i}$) before a contingency, and change in power injections ($\Delta P_{gen,i},\Delta P_{load,i},\Delta Q_{load,i}$) post contingency
    \item edge feature $\bm{x}_{s,t}$: line admittance and shunt parameters of the transmission line, $\bm{x}_{s,t}=[G,B,B_{sh}]$
\end{itemize}
\subsection{Training the model with a surrogate loss}\label{sec:training}
\begin{figure}[h]
	\centering
	\includegraphics[width=0.8\linewidth]{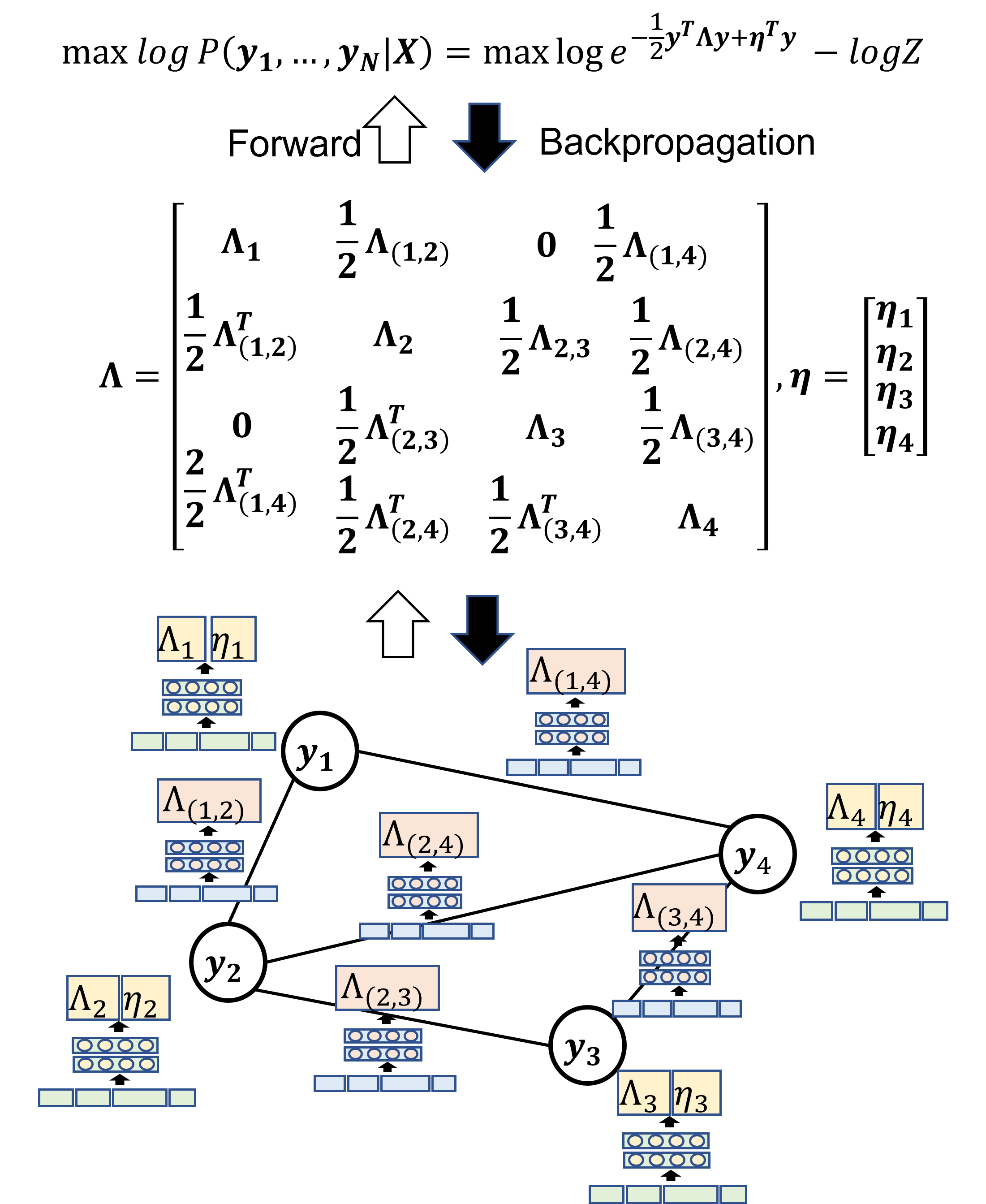}
	\caption[]{Training of the proposed method: forward pass and back-propagation.}
	\label{fig: training process}
\end{figure}

With the conditional GRF model defined in Section \ref{sec:cGRF} and the NN models designed in Section \ref{sec: NN-node and NN-edge}, the training process is illustrated in Fig. \ref{fig: training process}, where the forward pass of NN-node and NN-edge gives $\bm{\Lambda,\eta}$, and then the loss defined from the cGRF can be calculated to further enable a backward pass that updates the parameter $\bm{\theta}$.

As described in (\ref{eq: loglikelihood})-(\ref{eq:MLE simplified}), the loss function is the negative log-likelihood loss over the training data. Making use of the nice properties of Gaussian distribution, the partition function $Z(\bm{x,\theta})$ in the loss can be calculated analytically:
\begin{align}
    Z &= \int_{\bm{y}} exp(\bm{\eta}^T\bm{y}-\frac{1}{2}\bm{y}^T\Lambda\bm{y}) d\bm{y}\notag\\
&=\sqrt{\frac{2\pi}{|\Lambda|}} exp({\frac{\mu^T\Lambda\mu}{2}})
=\sqrt{\frac{2\pi}{|\Lambda|}} exp({\frac{\eta^T\Lambda^{-1}\eta}{2}})
    \label{eq: Z}
\end{align}

\noindent The detailed derivation is documented in Appendix \ref{appendix: Z}.

Furthermore, to enable a valid distribution $P(\bm{y}|\bm{x},\bm{\theta})$ and unique solution during inference, it is required that the $\bm{\Lambda}$ matrix is positive definite (PD), i.e., $\bm{\Lambda}\succ \bm{0}$. Therefore, adding this constraint and substituting (\ref{eq: Z}) into the loss function, the optimization problem of the proposed method can be written as:
\begin{align}
    &\min_{\bm{\theta}}\sum_{j=1}^{N}
    \frac{1}{2}\bm{y^{(j)T}}\bm{\Lambda^{(j)}}\bm{y^{(j)}}\notag\\
    &-\bm{\eta^{(j)T}}\bm{y^{(j)}}
    -\frac{1}{2}log|\bm{\Lambda^{(j)}}| 
    + \frac{1}{2}\bm{\eta^{(j)T}\Lambda^{-1(j)}\eta^{(j)}}\\
    s.t. &\notag\\
    &\text{(forward pass) }\bm{\Lambda^{(j)}}=f_\Lambda(\bm{x^{(j)}},\bm{{\theta_{\Lambda}}}),\forall j\\
    &\text{(forward pass) } \bm{\eta^{(j)}}= f_\eta(\bm{x^{(j)}},\bm{{\theta_{\eta}}}), \forall j\\
    &\text{(positive definiteness) }\bm{\Lambda^{(j)}}\succ \bm{0}, \forall j
    \label{eq: actual optimization problem}
\end{align}

In this problem, maintaining the positive definiteness of matrix $\bm{\Lambda}$ for every sample is required, not only in the final solution (i.e., throughout the training process due to the $log|\bm{\Lambda}|$ in the loss). This can be computationally challenging, especially when the network size grows with large matrix dimensions. 

To address this issue, we design a \textbf{surrogate loss} $\sum_{j=1}^{N}\frac{1}{2}\bm{(y^{(j)}-\mu^{(j)})}^T(\bm{y^{(j)}-\mu^{(j)}})$, which acts as a proxy for the actual loss we want to minimize. With the use of surrogate loss function, the  overall problem converts into the following form:
\begin{align}
    &\min_{\bm{\theta}}\sum_{j=1}^{N}\frac{1}{2}\bm{(y^{(j)}-\mu^{(j)})}^T(\bm{y^{(j)}-\mu^{(j)}})\\
    s.t. &\notag\\
    &\text{(forward pass) }\bm{\Lambda^{(j)}}=f_\Lambda(\bm{x^{(j)}},\bm{{\theta_{\Lambda}}}),\forall j\\
    &\text{(forward pass) } \bm{\eta^{(j)}}= f_\eta(\bm{x^{(j)}},\bm{{\theta_{\eta}}}), \forall j\\
    &\text{(inference) } \bm{\mu^{(j)}=\Lambda^{-1(j)}\eta^{(j)}}, \forall j
    \label{eq: surrogate optimization}
\end{align}

\noindent Appendix \ref{appendix: surrogate loss} shows how the new loss surrogate mathematically approximates the original objective function. In this way, we removed the need to maintain positive-definiteness of the $\bm{\Lambda}$ in the learning process, whereas the prediction is made using a $\bm{\Lambda}$ computed from the forward pass, and thus it still considers the power grid structure enforced by the graphical model. 

From decision theory, both the original and the surrogate loss aim to return an optimized model whose prediction $\bm{\hat{y}}$ approximates the ground truth $\bm{y}$, and both make predictions by finding out the linear approximation of the post-contingency system $\hat{\bm{\Lambda}}\bm{\hat{y}}=\hat{\bm{\eta}}$. 

Additionally, this surrogate optimization model can be considered as minimizing the mean squared error (MSE) loss $\frac{1}{2}||\bm{y-\hat{y}}||^2$ over the training data, where $\bm{\hat{y}}$ is the prediction (inference) made after a forward pass. 

\section{Incorporating More Physics}
\subsection{Parameter sharing: a powerful regularizer}
With each node having its own \textit{NN-node} and each edge having its \textit{NN-edge}, the number of parameters grows approximately linearly with grid size (more specifically, the number of nodes and edges). Can we reduce the model size further? The answer is yes! One option is to make all nodes share the same \textit{NN-node} and all edges share the same \textit{NN-edge}, so there are only two NNs in total.

Why does this work? Such sharing of \textit{NN-node} and \textit{NN-edge} is an extensive use of \textbf{parameter sharing} to incorporate domain knowledge into the network. Especially, from the physical perspective:

\noindent \textbf{Domain knowledge of location-invariant (LI) properties:} \textit{the power grid and the impact of its contingencies have properties that are invariant to change of locations: 1) any location far enough from the contingency location will experience little local change. 2) change in any location will be governed by the same mechanism, i.e., the system equations and Kirchhoff's laws.}

%Therefore, the same mapping can be used at all locations to learn the shared mechanism, with relevant neighborhood features fed as input. 
Parameter sharing across the grid network significantly lowers the number of unique model parameters. Also, it reduces the need for a large amount of training data to adequately learn the system mapping for larger grid sizes (like the networks representing continental U.S. networks with $>80k$ nodes).

\subsection{Zero-injection bus}
\noindent \textbf{Domain knowledge of zero-injection (ZI) buses:} \textit{ A bus with no connected generation or load is called a zero-injection (ZI) bus. These buses neither consume nor produce power, and thus, injections at these buses are zero.
}

%A realistic and feasible solution around ZI buses should guarantee that the total output current on its branches sum up to $0$. 
In the proposed approach, the model parameter $\eta$ serves as a proxy to bus injections; therefore, we can integrate domain knowledge about zero-injection nodes into the method by setting $\bm{\eta}_i=0$ at any ZI node $i$.

% \subsection{Incorporating pre-contingency solution as prior knowledge} #bad performance

% \subsection{Using pre-contingency voltage as a physics-informed prior} \reminder{may not needed}

% \noindent \textit{\textbf{(Domain knowledge:)}
% Empirically, we can notice that when a contingency happens at a certain location, the bus states nearby the contingency location are most impacted, whereas far-away buses remain close to the pre-contingency state. 
% }

% This domain knowledge could be used in model design, through a usage of prior distribution on output $y$. By applying this domain knowledge, we can improve the accuracy of predictions on locations far away from the contingency and improve the generalization of the method.

\section{Experiments}\label{sec:experiments}
\label{sec:Results}
This section runs experiments for CA instances in the context of MadIoT attack. The experiments are to validate that the proposed warm starter provides good initial states for \textit{hard-to-solve} N-k contingencies and enables faster convergence when compared to traditional initialization techniques.

We test three versions of the proposed method. These versions differ in the level of domain-knowledge that they incorporate within their model. Table \ref{tab:summarize knowledge} summarizes the domain-knowledge in these versions. Table \ref{tab:summarize methods} categorizes the domain knowledge in each version.

\setlength{\tabcolsep}{6pt}
\begin{table}[htbp]
\small
\centering
	\caption{Summary of domain knowledge \label{tab:summarize knowledge}}
	\begin{tabular}{ @{}rll@{} }  
	\toprule
	\textbf{Knowledge} & \textbf{Technique} &\textbf{Benefits} \\ \midrule
		\textbf{topology}  
     & graphical model & - physical interpretability\\
     &  &- generalization (to topology)\\
     \midrule
     \textbf{decisive}
     & feature selection  & - accuracy \\
      \textbf{features}
     & & - physical interpretability\\
     & & - generalization (to load\&gen)\\
     \midrule
     \textbf{taylor}
     &feature selection & - accuracy \\
     \textbf{expansion}
     & & - physical interpretability\\
     \midrule
     \textbf{LI properties}
     &parameter & - trainability, scalability \\
     &sharing (PS) &- generalization ($\downarrow$ overfitting)\\
     \midrule
     \textbf{ZI bus}
     &enforce $\eta_i=0$ & - physical interpretability \\
     & & - generalization \\
	\bottomrule
	\end{tabular} 
\end{table}

\setlength{\tabcolsep}{6pt}
\begin{table}[htbp]
\small
\centering
	\caption{Categorization of domain-knowledge in the 3 versions \label{tab:summarize methods}}
	\begin{tabular}{ @{}rccc@{} }  
	\toprule
	\textbf{Knowledge \&} & \textbf{cGRF} & \textbf{cGRF-PS} & \textbf{cGRF-PS-ZI}\\ 
	\textbf{techniques} &  & & \\
	\midrule
	graphical model (cGRF) & \checkmark& \checkmark& \checkmark\\	
	\midrule
	feature selection & \checkmark&  \checkmark& \checkmark\\
	\midrule
	parameter sharing (PS) & & \checkmark& \checkmark\\
	\midrule
	ZI buses & & &\checkmark\\
	\bottomrule
	\end{tabular} 
\end{table}

\subsection{Data generation and experiment settings}
We generate synthetic MadIoT contingencies for the following two networks: i) IEEE 118 bus network \cite{IEEE118} ii) synthetic Texas ACTIVSg2000 network \cite{ACTIVSg2000}. {\color{black}For each network, we generate $N_{data}$ contingency samples $\{(x^{(j)},y^{(j))}\}_{N_{data}}$ where feature the notation of $x$ and $y$ has been illustrated in Fig. \ref{fig: graphical model} and Table \ref{tab:notations}. } The algorithm to generate the synthetic contingency data is given in Algorithm \ref{alg: main generation}.
\begin{algorithm}
	\caption{3-Step Data Generation Process} 
	\label{alg: main generation}
	\KwIn{Base case $G_{base}$, type of contingency $t_c$, number of data samples $N_{data}$}
	\KwOut{Generated dataset $\{(x^{(j)},y^{(j))}\}_{N_{data}}$}
	\For{{$j \gets 1$ to $N_{data}$} }{
	{\bf 1. Create a random feasible pre-contingency case $G_{pre}^{(j)}$:} each sample has random topology, generation and load level. 
	
	{\bf 2. Create contingency $c^{(j)}$ on $G_{pre}^{(j)}$: } which has attributes \textit{type, location, parameter}.
	%e.g.,$$c^{(j)}=\{'type':MadIoT,'location'=[2,3],'parameter'=1.2\}$$ denotes a MadIoT attack scenario that scales up the loads on bus# $2$ and $5$ to $120\%$ the original load values.
	
	{\bf 3. Simulate with droop control:} run power flow to obtain the post-contingency voltages $\bm{v}^{post}$
	}
\end{algorithm}

{\color{black}\textbf{Contingency set and model generalization:} In our experiment, we train and test warm starter for a MadIoT scenario of  increasing the top K largest loads by the same amount (percentage) which is a severe scenario threatening the power grid. And the pre-contingency load is randomly sampled within the range of 95\%-105\% base load, and topology is randonly sampled by disconnecting 1-2 random lines on the base case, to represent the different normal operating conditions of a power system. Such contingency generation settings to a very specific setup of the dataset so that learning becomes more targeted: these contingencies are ”hard-to-solve” for an optimization solver, but a simple and interpretable learned
model might be able to easily extract the major relationships to provide good warm-start values. 
Obviously, this leads to limited generalization issue that the trained model can hardly apply to a different contingency scenario where loads are manipulated on other locations and by another amount.
But in practice, this can be addressed with multiple models. The operators or decision makers can decide several other significant contingency settings that are worth consideration and evaluation. And they can train a second model on a second dataset which describes another important contingency setting. So that the different severe cyberthreat scenarios can be considered with learning performed in a targeted way.  
}

The experiment settings that are used for the data generation, model design, and model training are documented in Table \ref{tab:experiment settings}. 
{\color{black} Based on the idea of \textit{NN-node} and \textit{NN-edge} in Section \ref{sec: NN-node and NN-edge}, the neural networks used in this work aim to learn a low-dimensional mapping from local node / edge input features to local outputs to form $\Lambda, \eta$. This can be effectively done with simple and shallow neural network architectures. In our experiment, the model is designed with a shallow 3-layer NN architecture with 64 hidden layers in each, to save computation time and reduce overfitting. It also allows us to experiment on whether a simple model design can give good performance.
The training is then done with an Adam optimizer and step learning rate scheduler.}

\setlength{\tabcolsep}{6pt}
\begin{table}[htbp]
\small
\centering
	\caption{Experiment settings \label{tab:experiment settings}}
	\begin{tabular}{ @{}rl@{} }  
	\toprule
	\textbf{Settings} & \textbf{(see Table \ref{tab:notations} for definitions)} \\ \midrule
     $N_{data}$ & 
     case118: 1,000; ACTIVSg2000: 5,000  \\
     & split into train, val, test set by $8:1:1$\\
     \midrule
     \textit{NN-node} \& \textit{NN-edge}& shallow cylinder architecture\\
     &$(n\_layer,hidden\_ dim)=(3,64)$\\
     \midrule
     contingency $c$ & $type:$ MadIoT\\
     & $location:$ randomly sampled 50\% loads\\
     & $parameter:$ case118 200\%,\\ &  ACTIVSg2000 120\%\\
     \midrule
     optimizer & $Adam,lr=0.001, scheduler=stepLR$\\
	\bottomrule
	\end{tabular} 
\end{table}

\subsection{Physical Interpretability}
\begin{figure}[h]
	\centering
	\includegraphics[width=0.75\linewidth]{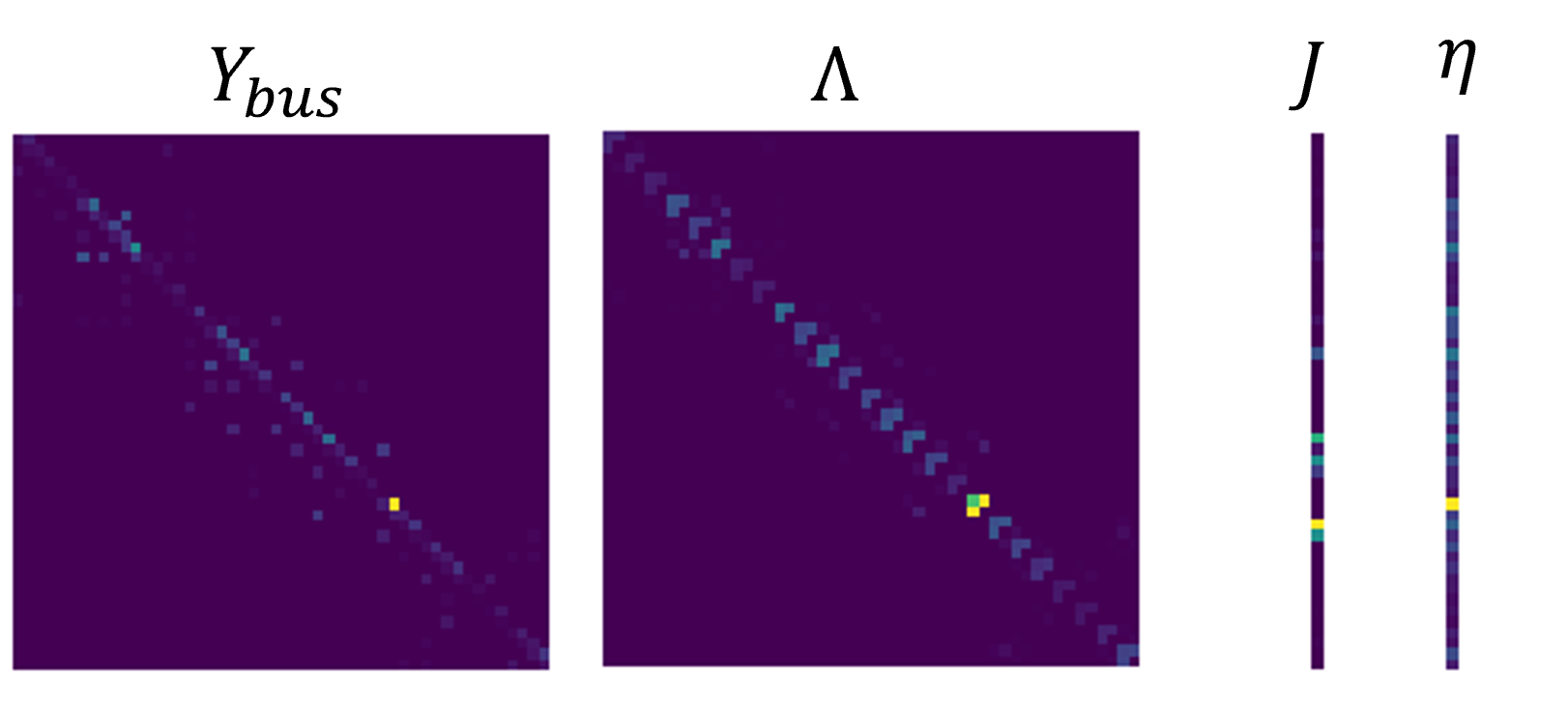}
	\caption[]{Physical interpretation: \color{black}This figure visualizes the values in matrices and vectors (the more yellow, the larger the value). Learned model parameters $\Lambda, \eta$ have some similarity with the true post-contingency system admittance matrix $Y_{bus}$ and injection current vector $J$, in terms of the sparse structure and value distribution. This is because the learned parameters $\Lambda, \eta$ aims to form a linear model $\Lambda y= \eta$ which is a linear proxy of the true linearized system model $Y_{bus}y=J$.}
	\label{fig: interpretation}
\end{figure}

Fig. \ref{fig: interpretation} validates our hypothesis in Section \ref{sec:cGRF} about \textit{the physical interpretation of our method as a linear system proxy}. In Fig. \ref{fig: interpretation}, we visualize the result on a test sample to show the similarity between the linear proxy given by model parameters $\Lambda, \eta$ and the true post-contingency system linearized admittance matrix $Y_{bus}$ and injection current vector $J$ at the solution; thus, validating that the model acts as a linear proxy for the post-contingency operating condition.

\subsection{Application-level practicality}

To verify the effectiveness of the warm starter, we compare the convergence speed ($\#$ iterations) with three different initialization methods for the physical solver \cite{sugar-pf}: 
\begin{enumerate}
    \item flat start (flat)
    \item pre-contingency solution ($V_{pre}$)
    \item physical solver warm-started by the three versions of the proposed method (cGRF)
\end{enumerate}

%In case when the physical solver is used without the cGRF warm starter, to ensure robust solution methodology, we apply homotopy [xx] when the standard method fails. We consider the physical solver to have failed when it cannot converge within a predefined number of iterations or the residual values blow up. 
\begin{figure*}[h]
	\centering
	\includegraphics[width=0.9\linewidth]{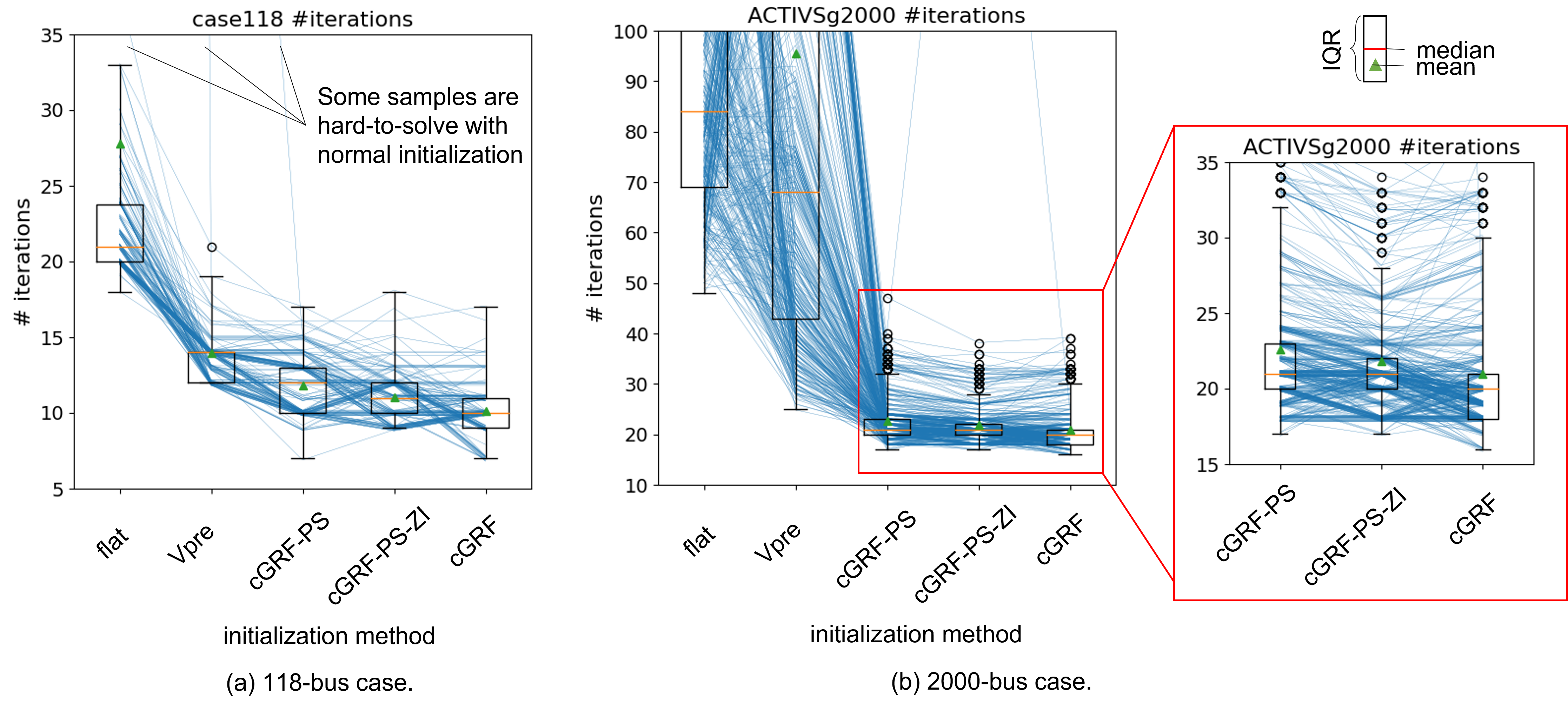}
	\caption[]{Result on test data: power flow simulation takes fewer iterations to converge with the proposed method, than traditional initialization methods:  i. \textbf{flat} start: starting with $(V_i, \delta_i) \leftarrow (1,0), \forall i \in \{1,...,n\}$; ii). \textbf{Vpre} start: warm-starting from pre-contingency voltages.}
	\label{fig:iterations case118}
\end{figure*}

Fig. \ref{fig:iterations case118} shows the evaluation results on test data. These include the test samples  {\color{black}  (we split the train, validate, and test set by 8:1:1 as illustrated in Table \ref{tab:experiment settings}) which include 100 unseen contingency samples for case 118, and 500 unseen contingency samples} for ACTIVSg2000.  For cGRF results, we feed the ML predictions into a power flow simulator: SUGAR \cite{sugar-pf}. The simulator is robust because it always converges. In case the general NR loop fails, the simulator uses homotopy \cite{sugar-pf} to ensure convergence, albeit at a computational cost.

The results show that simulation takes fewer iterations to converge with the proposed ML-based cGRF initialization when compared to traditional initialization methods (flat or $V_{pre}$). In particular, on ACTIVSg2000, many contingency samples are hard-to-solve with traditional initialization (and may require the homotopy option in SUGAR). In contrast, our ML-based cGRF method significantly speeds up convergence (up to 5x improvement in speed) even with the shallow 3-layer NN architecture.

Moreover, due to the use of parameter sharing (PS) which enables NNs to share parameters, the \textit{lightweight} model cGRF-PS significantly reduces the total number of model parameters but achieves comparable results to the base model cGRF. In particular, while the average doesn't decrease significantly, the variation decreases due to the use of zero injection (ZI) knowledge. And cGRF-PS-ZI further shows that integrating more grid physics into the model through zero injection (ZI) knowledge can further improve convergence of the lightweight model.

\section{Conclusion}
\label{sec:Conclusion}
This paper proposes a novel physics-informed ML-based warm starter for cyberthreat-focused contingency analysis. Our method has the following features:
    \begin{itemize}
        \item \textbf{generalizability to topology changes} by using a graphical model to naturally represent grid structure
        \item \textbf{physically interpretability} by generating 'global' predictions from a system-level linear proxy
        \item  \textbf{scalability} by using the graphical model and parameter sharing techniques
    \end{itemize}
While being generic in its approach, the method is designed to speed up the simulation of $N-k$ contingency events. We believe these contingencies will be included in the future to evaluate the grid's vulnerability to cyberthreat instances, such as those from the MadIoT attack. In the results, we show that the proposed method can reduce the simulation iteration count by up to 5x, when compared with traditional initialization methods for such contingencies.

\section*{Acknowledgment}
Work in this paper is supported in part by C3.ai Inc. and Microsoft Corporation.
%The preferred spelling of the word ``acknowledgment'' in America is without 
%an ``e'' after the ``g''. Avoid the stilted expression ``one of us (R. B. 
%G.) thanks $\ldots$''. Instead, try ``R. B. G. thanks$\ldots$''. Put sponsor 
%acknowledgments in the unnumbered footnote on the first page.
\bibliographystyle{IEEEtran}
\bibliography{pscc2024}

\section{Appendix}
%\subsection{Markov properties and physical meanings}\label{appendix:markov properties}

% \subsection{Why post-contingency voltage can be modeled as a distribution?}\label{appendix:v distribution}
% There can be several ways to interpret
% \begin{enumerate}
%     \item The output solution $y$ in the dataset can have small errors from the true behavior of a system or the optimal solution of an optimal simulator. Modeling it as a distribution assumes that the dataset approximates the ground truth with some precision.
%     \item The simulation results $y$ for the same input feature $x$ using different solvers may vary. Such variance within the dataset can be considered by the distribution.
%     \item As our warm-starter is used for a single point estimation, rather than statistical distribution of the output, $P(\bm{y}|\bm{x},\bm{\theta})$ can theoretically be any distribution whose expectation approximates the desired output to some precision. It need not have real physical meaning.
% \end{enumerate}

\subsection{Calculate partition function  }\label{appendix: Z}
For a multivariate Gaussian distribution $\bm{y}\sim N(\bm{\mu},\bm{\Sigma})$ where $\bm{\mu}$ denotes the mean and $\bm{\Sigma}$ denotes the covariance matrix, let $\bm{\Lambda}=\bm{\Sigma}^{-1}$, we have:
\begin{equation}
   \int_{\bm{y}} \sqrt{\frac{|\bm{\Lambda}|}{2\pi}}exp({-\frac{1}{2}(\bm{y-\mu})^T\bm{\Lambda}(\bm{y-\mu})}) d\bm{y}= 1
   \label{eq: standard Gaussian pdf integral}
\end{equation}

As mentioned earlier, the Gaussian CRF model $P(\bm{y}|\bm{x},\bm{\theta})=\frac{1}{Z(\bm{x,\theta})} exp(\bm{\eta}^T\bm{y}-\frac{1}{2}\bm{y}^T\Lambda\bm{y})$
is equivalent to a multivariate Gaussian distribution $N(\bm{\mu},\bm{\Sigma})$ with
$\bm{\eta}=\bm{\Lambda}\bm{\mu}, \bm{\Lambda}=\bm{\Sigma}^{-1}$. Thus (\ref{eq: standard Gaussian pdf integral}) can be rewritten as:
\begin{equation}
    \sqrt{\frac{|\bm{\Lambda}|}{2\pi}}exp({-\frac{\mu^T\Lambda\mu}{2}})\int_{\bm{y}} exp(\bm{\eta}^T\bm{y}-\frac{1}{2}\bm{y}^T\Lambda\bm{y}) d\bm{y}= 1
\end{equation}

Taking the nice properties of Gaussian distribution, the partition function $Z(\bm{x,\theta})$ can be calculated as:
\begin{equation}
   Z(\bm{x,\theta}) = \int_{\bm{y}} exp(\bm{\eta}^T\bm{y}-\frac{1}{2}\bm{y}^T\Lambda\bm{y}) d\bm{y}
=\sqrt{\frac{2\pi}{|\Lambda|}} exp({\frac{\mu^T\Lambda\mu}{2}})
\end{equation}

\subsection{Surrogate loss }\label{appendix: surrogate loss}

Mathematically, due to the Gaussian distribution properties, the original optimization problem in (\ref{eq: actual optimization problem}) is equivalently:
\begin{align}
    &\min_{\bm{\theta}}\sum_{j=1}^{N}\frac{1}{2}\bm{(y^{(j)}-\mu^{(j)})}^T\bm{\Lambda^{(j)}}(\bm{y^{(j)}-\mu^{(j)}})
    -\frac{1}{2}log|\bm{\Lambda^{(j)}}| \\
    s.t. &\notag\\
    &\text{(forward pass) }\bm{\Lambda^{(j)}}=f_\Lambda(\bm{x^{(j)}},\bm{{\theta_{\Lambda}}}),\forall j\\
    &\text{(forward pass) } \bm{\eta^{(j)}}= f_\eta(\bm{x^{(j)}},\bm{{\theta_{\eta}}}), \forall j\\
    &\text{(positive definiteness) }\bm{\Lambda^{(j)}}\succ \bm{0}, \forall j\\
     &\text{(inference) } \bm{\mu^{(j)}=\Lambda^{-1(j)}\eta^{(j)}}, \forall j
\end{align}

To design a surrogate loss, we make an approximation $\bm{\Lambda^{(j)}=I}$ ($\bm{I}$ is identity matrix) only in the objective function, so that $log|\bm{\Lambda}|=0$ becomes negligible.

\end{document}